\newcommand{\HI}{H{\small\,I}}
\newcommand{\HII}{H{\small\,II}}
\newcommand{\Halpha}{H$\alpha$}
\newcommand{\Hbeta}{H$\beta$}

\newcommand{\kms}{$\,$km$\,$s$^{-1}$}

\newcommand{\Msol}{{M$_\odot$}}

\documentstyle[11pt,paspconf]{article}
\input psfig

\begin{document}

\title{\HI\ in Low-Luminosity Early-Type Galaxies}

\author{Tom Oosterloo}
\affil{CNR - Istituto di Fisica Cosmica, Milan, Italy}

\author{Raffaella Morganti}
\affil{CNR - Istituto di Radioastronomia, Bologna, Italy}

\author{Elaine Sadler}
\affil{Sydney University, Australia}

\begin{abstract}
We discuss the properties of the \HI\ in low-luminosity early-type galaxies.
The morphology of the \HI\ is more regular than that of the \HI\ in many
more-luminous early-type galaxies.  The \HI\ is always distributed in a disk
and is more centrally concentrated. The central \HI\ surface densities are
higher than in luminous early-type galaxies and are high enough for star
formation to occur.

\end{abstract}


\section{Why study H{\small\bf\,I} in low-luminosity early-type galaxies}

A few decades  ago, elliptical galaxies appeared to  form a homogeneous group.
We know now that in reality  they constitute a  very varied family of galaxies
of which many of the detailed characteristics vary from galaxy to galaxy.  For
many of these  characteristics,  elliptical galaxies can  be  divided into two
groups, of which we list a few.

{\sl Stellar   rotation vs anisotropy.}  Some  elliptical galaxies  are purely
'pressure supported'  systems, i.e.\ the  random motions  of the stars support
the  shape of   the galaxy, while   for others  the  rotation  of the  stellar
component  is  important  for understanding  the   dynamical structure of  the
galaxy.

{\sl Boxy vs   Disky.} For many ellipticals,  the  isophotes  are not  perfect
ellipses.  For some, the deviations are boxy-shaped, while for others they are
disky.  For the disky  galaxies, there is  a continuum in bulge-to-disk ratio,
from genuine spiral  galaxies all the   way to 'no  disk'  (e.g., Kormendy and
Bender, 1996).

{\sl Core properties.}  Imaging the centres of  elliptical galaxies with  {\sl
HST} has shown that the  density distribution in  the centre of ellipticals is
quite different from galaxy  to   galaxy.  Some ellipticals  have   relatively
shallow central cusps, while others show a steep cusp.  The shape of this cusp
appears to depend on the luminosity of the galaxy (e.g., Lauer 1997).

{\sl AGN.} The radio continuum emission from the nuclei of elliptical galaxies
varies strongly from  galaxy to galaxy.    The fraction of ellipticals with  a
radio core  depends strongly on the  luminosity, although for  a given optical
luminosity there is still a large scatter ($\sim$4 orders of magnitude) in the
radio power (Sadler et al.\ 1989).

{\sl X-ray gas.} The amount of X-ray emission correlates strongly with optical
luminosity.  In large ellipticals, part  of the X-ray emission originates from
a halo of hot gas, while in smaller ellipticals the X-ray is due only to X-ray
binaries (e.g., Canizares et al.\ 1987).

{\sl  Liner vs H{\small\sl\,II}    spectrum.}  Many ellipticals show   optical
emission from ionized gas.  The character of the  spectrum of this gas changes
with luminosity.  The  smaller galaxies show a  \HII\ region-like spectrum and
hence the emission is likely to be due to star formation.  In the more massive
galaxies, the character changes to  that of a  Liner spectrum, indicating that
the ionization mechanism is different in these galaxies (e.g., Sadler 1987).

{\sl Star  formation  history.} The  star  formation history   differs between
galaxies.    For example, the   relative abundance  of   Mg with respect to Fe
correlates with velocity dispersion.  Luminous  ellipticals have [Mg/Fe] up to
0.4, while fainter ellipticals have values around zero.  This could point to a
different star  formation history,  with differences   in the  way the ISM  is
enriched.   Low-luminosity  ellipticals also   show a   larger spread  in  the
Mg-$\sigma$ relation, again pointing to  a different star formation history in
some of  these galaxies (e.g., Bender  1996).  Many low-luminosity ellipticals
in  fact display star formation   in the central  parts of  the galaxies.   It
appears   that disky galaxies  have  stronger \Hbeta\ indices, indicating that
some star formation occurred recently (de Jong \& Davies 1997).

Many  of these  differences  between  different galaxies can   be explained by
different   amounts  of  gas    (and   hence  dissipation)  present    in  the
formation/evolution, and in many models for galaxy formation the gas supply is
a key factor (e.g., Kauffmann 1996). For example, the differences between boxy
and disky galaxies, the importance of  rotation vs.\ anisotropic galaxies, and
the different central  density distributions can all  be a consequence  of the
relative importance  of gas.    Obviously, since stars  form from gas, the
different star formation histories must be related to  different gas contents
during the evolution.

From the above it follows that, to understand the reasons for the differences
between galaxies, one has to know about the gas.  Of course, most ellipticals
formed most of their stars a long time ago and we cannot see this gas directly,
but knowing about the current situation can still provide useful information. 
Since many characteristics appear to change systematically with mass, one would
like to have a picture of the gas properties as function of mass in
ellipticals.  Some data (i.e.\ full data-cubes of the \HI) are available for
lower-luminosity ellipticals (e.g.\ Lake et al.\ 1987; with low-luminosity we
mean galaxies with absolute magnitude in the range --16 to --19), but most of
the data published are of higher-mass galaxies (and even for those galaxies the
number of good datacubes is small).  Since it appears that for lower-luminosity
galaxies the gas was more important than for more massive galaxies, we need to
have as many as possible low-luminosity, early-type galaxies well studied in
\HI.

\begin{table}
\begin{center}
\begin{tabular}{lccc} 
\tableline 
Galaxy Type & Observed & Detected & \% \\
\tableline 
 E               &    64 & 3   &  5  \\
 E/S0            &    23 & 4   & 17  \\
 S0, SB0         &   103 & 21  & 20  \\
 Pec E and S0    &    20 & 9   & 45  \\
 S0/a and pec    &    35 & 15  & 43  \\
 Sa and pec      &   103 & 78  & 76  \\ 
\tableline 
\tableline 
\end{tabular}
\end{center}
\caption{\HI\ detection rates for early-type galaxies (from Bregman et al.\
1992}
\end{table}

In the past few years, we have been imaging the \HI\ in a number of elliptical
galaxies with the Australia telescope Compact Array (ATCA) (see also Morganti
et al.\ these proceedings).  Four of the galaxies we have observed (NGC 802,
NGC 2328, ESO 118--G34 and ESO 027--G21) are of lower luminosity and here we
present some of the results from these observations.  They are all early-type
galaxies with absolute magnitude between --18 and --19 (for $H_\circ = 50$
\kms\,Mpc$^{-1}$). 

\section{Optical Morphology}

When discussing the \HI\ content of early-type galaxies (or related issues like
star formation), it is important to stress which early-type galaxies are
considered.  It is, therefore, important to first address the issue of galaxy
morphology.  Unlike for spiral galaxies, the \HI\ content of early-type
galaxies varies greatly from galaxy to galaxy and galaxy morphology and
environment are key factors.  Hence, if one studies the morphology and the
kinematics of the \HI\ in early-type galaxies, one discusses a particular
subset of early-type galaxies, with biases in evolution state and environment. 
This is illustrated in  Table 1 that gives the \HI\ detection rates
for different types of early-type galaxies. 

Table 1 shows two things.  If elliptical galaxies are defined as
pressure-supported systems with no obvious peculiarities in their morphology,
the discussion on their \HI\ can be quite short: only 5 percent of these
galaxies are detected.  However, if galaxies with some peculiarity in their
optical morphology (like shells, dust lanes etc.) are also considered, the
detection rate increases dramatically.  This means that \HI-selected early-type
galaxies are a certain subsample of the whole population, namely those that
have had some interaction recently.  This indicates that the presence of \HI\
depends on certain aspects of the evolution of elliptical galaxies.  This is
important to keep in mind.  To get a complete picture of the evolution of
early-type galaxies, it is essential to include these \HI-rich galaxies, and
not restrict the samples to `pure' ellipticals with no optical peculiarities. 
The issue is to some extent whether we consider galaxies like Centaurus A to be
an elliptical or not.  Surely, many early-type galaxies have a similar
evolution history as Centaurus A and one cannot hope to understand the
evolution of early-type galaxies without considering galaxies like Centaurus A. 

The correlation between the \HI\ content and the presence of peculiarities
strongly suggests that in many early-type galaxies detected in \HI, the neutral
hydrogen is accreted or is left over from a merger (e.g.\ Knapp et al.\ 1985). 
But Table 1 suggests that perhaps in some galaxies this is not necessarily the
explanation for the presence of \HI.  Table 1 shows that the \HI\ content seems
to be related also to {\sl how much disk an early-type galaxy has}.  This is
also an important clue, certainly in the context of some of the characteristics
mentioned in the introduction (for example, the continuum from spirals to disky
ellipticals, or the star formation linked to disky-ness).

\begin{figure}
\centerline{\psfig{file=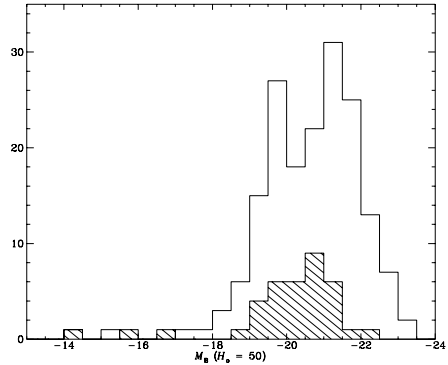,height=6cm}}
\caption{\HI\ detection rate of E and E/S0 galaxies as function of absolute
magnitude. The shaded histogram gives the distribution  of the detections, the
unshaded histogram of the non-detections.  Data from Bregman et al.\ 1992}
\end{figure}

\section{H{\small\bf\,I} content}

The  properties of low-luminosity  early-type galaxies  suggest  that gas  has
played a more important role in the formation/evolution of these galaxies than
in their  more   massive  counterparts.  Do  we   observe that  low-luminosity
early-type  galaxies have more   or more   often  \HI?  When one  reads  about
low-luminosity early-type galaxies, one often gets the impression that this is
the case. We think,  however, that the evidence  for this is not  very strong.
An  often quoted reference  is  Lake \& Schommer  (1984)  who observed a small
sample of  low-luminosity early-type galaxies.   The \HI\  detection rate they
obtained is significantly higher that those obtained in  other studies of more
massive elliptical galaxies available at the time.  The galaxies they observed
are however relatively nearby and the sample is not very  large, so it is not
clear how significant the result is.  Looking at larger samples available now,
there  is not much evidence that  low-luminosity  early-type galaxies are more
likely to contain \HI\

Figure 1 shows a histogram of the detection  rate of E  and E/S0 galaxies as a
function of absolute  magnitude.  This histogram does  not give  an indication
that early-type galaxies  with $M_B$ between --16  and --19 are more likely to
have \HI.  However, a problem is that the sample  is not entirely homogeneous.
Many of the galaxies  in this sample are in  the Virgo cluster  (because Virgo
passes straight over Arecibo) and Virgo  ellipticals are poor  in \HI\ (as are
the spirals in Virgo).  The `contamination' by Virgo is especially significant
for galaxies fainter  than $M_B \sim  -19$, so  if galaxies in  this magnitude
range are more likely to have \HI, this could be suppressed in the sample.

This raises  again  the issue  of  biases:  the  depletion  of \HI\  in  dense
environments  means that studying \HI-selected  early-type galaxies one biases
in favour  of  field galaxies,  and hence  for galaxies  that may  have  had a
different evolution compared to cluster ellipticals.

One interesting point is that it looks like that the largest ellipticals ($M_B
< -22$) are poorer in \HI.  This could be an environmental effect as well.

In summary, there is no evidence that intermediate luminosity galaxies are more
likely to have \HI, but the available sample is quite small and biased by
Virgo.  On the other hand, it is also quite unlikely that these galaxies are
poorer in \HI.  Several galaxies have now been detected and the \HI\ in these
galaxies has been imaged.

\begin{figure}
\centerline{\psfig{file=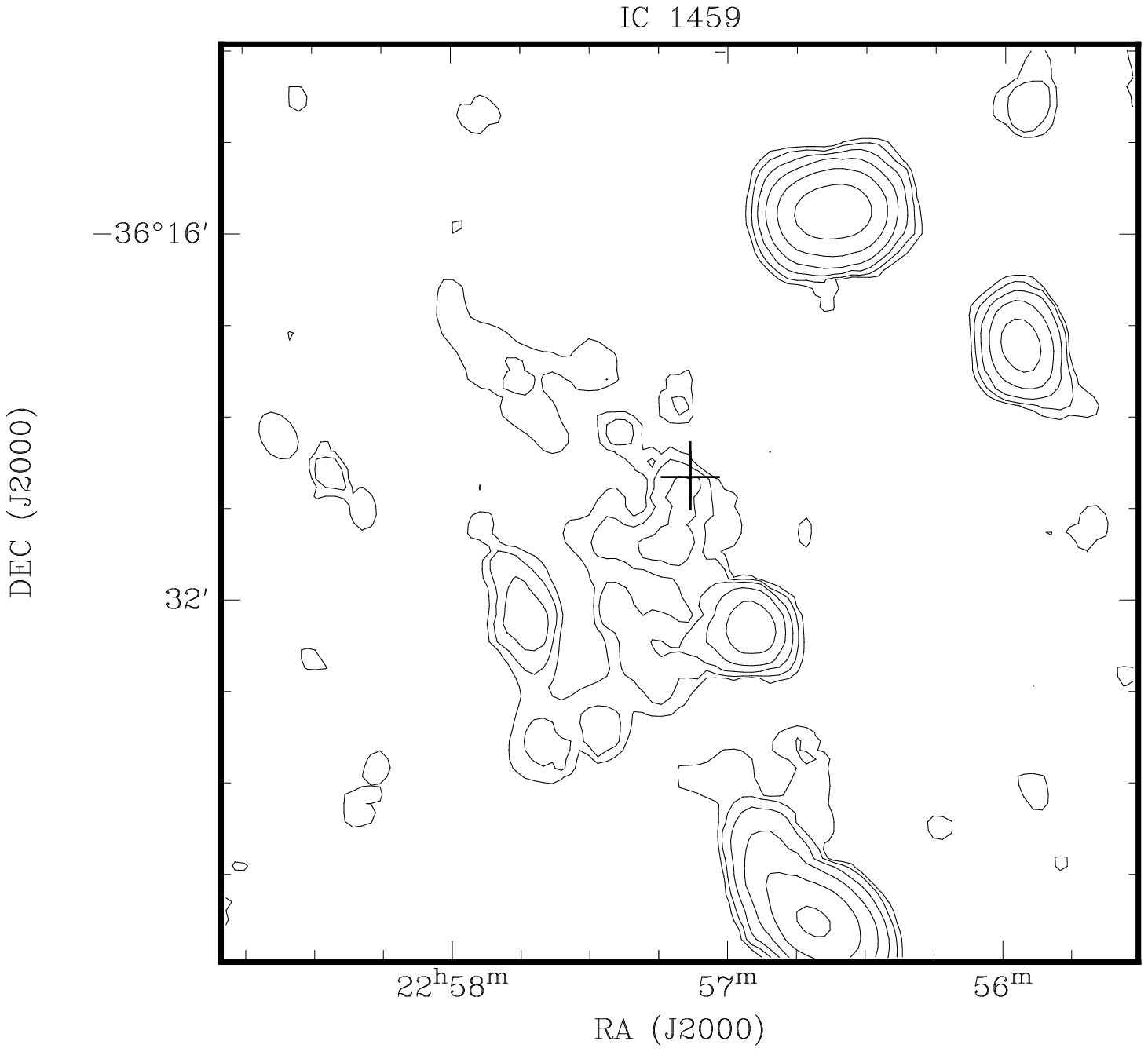,width=6.5cm}
\hss{\psfig{file=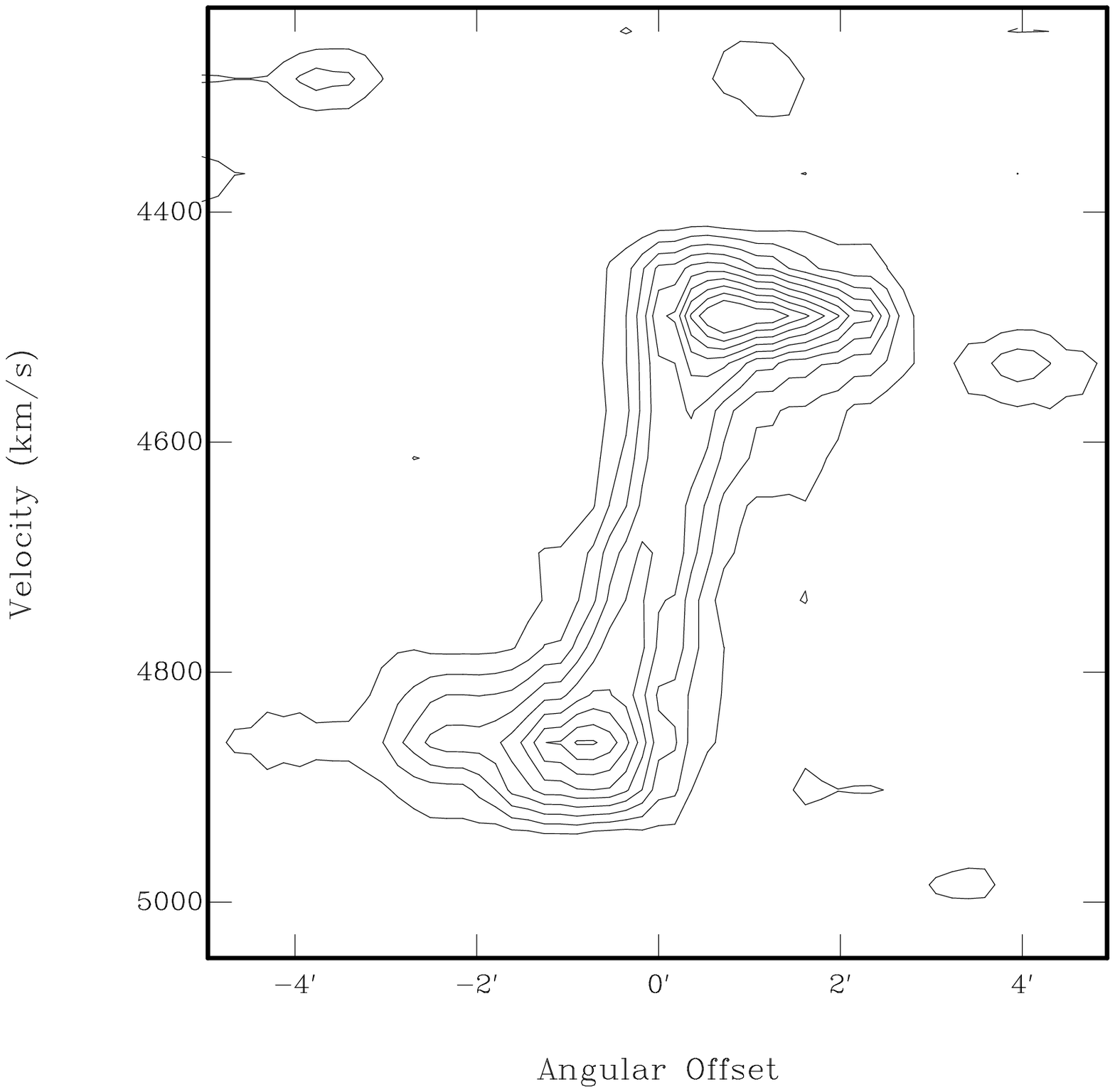,width=5.5cm}}}
\caption{{\sl left)} total \HI\ image of the field around IC 1459 (identified
by the cross near the centre). {\sl right)} Position-velocity plot of NGC 807,
taken along the kinematical major axis (data taken from the VLA archive,
original data taken by Dressel in 1985)}
\end{figure}

\section{H{\small\bf\,I} Morphology and Kinematics}

The main information about the morphology of the \HI\ in low-luminosity
galaxies comes from Lake et al.\ (1987) with 4 studied galaxies and our more
recent observations of another 4 galaxies with the ATCA. 

The morphology of the \HI\ in low-luminous early-type galaxies is different
from that in more luminous galaxies.  In the latter, the \HI\ morphology varies
very much from galaxy to galaxy, while in low-luminosity galaxies, the range in
morphologies is much smaller.  In many luminous ellipticals, the morphology of
the \HI\ is quite irregular and the \HI\ is evidently accreted.  One example is
the well-known galaxy IC 1459.  Figure 2a shows the total \HI\ image we
obtained for this galaxy with the ATCA.  All the bright \HI\ emission comes
from galaxies surrounding IC 1459, but there is also lower surface-brightness
\HI\ that is not clearly associated with any of these galaxies.  The \HI\
appears to be stripped from the \HI-rich galaxies, most likely due to
interactions between these galaxies and IC 1459.  Such irregular \HI\
structures around more luminous elliptical galaxies are quite common (other
examples are NGC 5077 and NGC 4936).  None of the low-luminosity galaxies that
have been observed show such \HI\ characteristics. 

In other luminous elliptical galaxies, the morphology of the \HI\ is clearly
associated with the galaxy, but the morphology of the \HI\ is often still
relatively irregular, although the velocity fields are more regular than
one would expect from the density distribution (e.g.\  Schiminovich and van
Gorkom 1997, Morganti et al.\ this proceedings). Usually, these galaxies show
other indications that the \HI\ is likely accreted (shells or dust lanes).

\begin{figure}
\centerline{\psfig{file=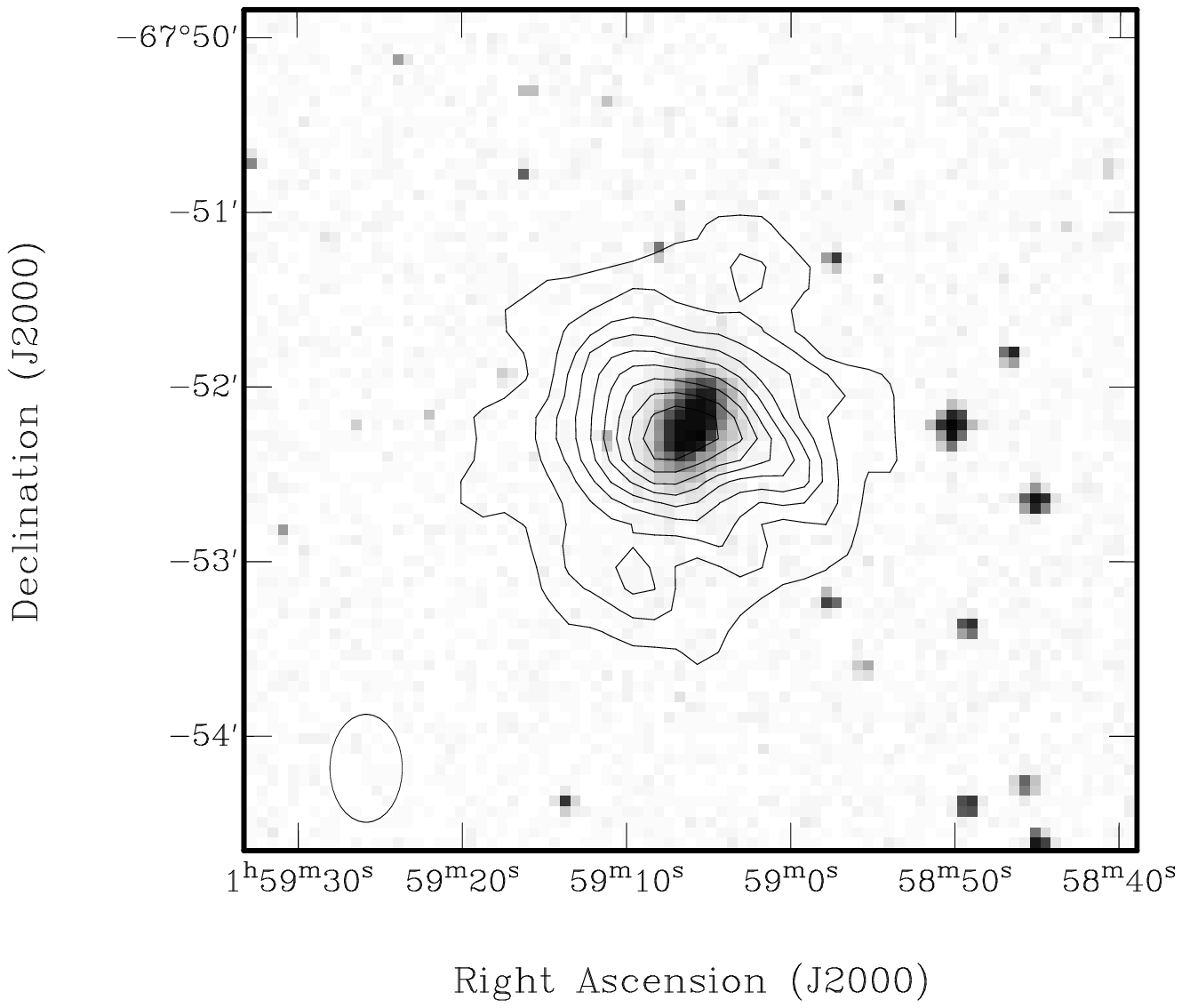,width=5cm}
\hss
\psfig{file=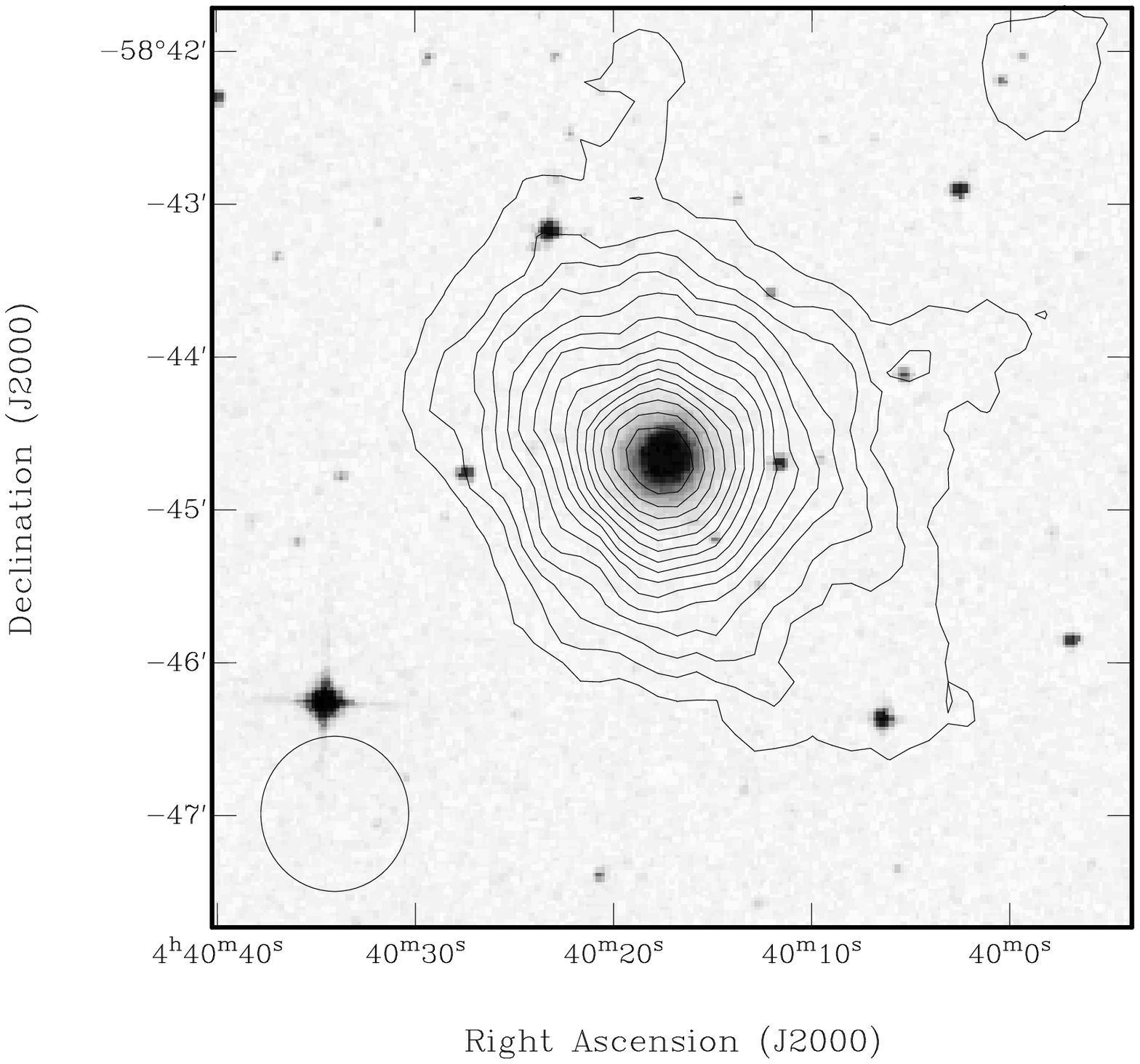,width=5cm}}
\caption{Total \HI\ images of NGC 802 and ESO 118--G34}
\end{figure}

However, there are a few reasonably luminous galaxies known now that have a
regular \HI\ disk, where there are no obvious signs, that the \HI\ is accreted
recently.  These gas disks could be the gas counterpart of a very faint stellar
disk in these systems.  One example of such a \HI\ disk is NGC 807.  Figure 2b
shows the position-velocity diagram (taken along the kinematical major axis) of
the \HI\ disk in this galaxy.  The difference between this \HI\ disk and those
in spiral galaxies is that the surface density of the \HI\ disk is a factor
5-10 lower in elliptical galaxies.  The peak surface densities are at most 2
\Msol\,pc$^{-2}$, even lower than observed in low surface-brightness spiral
galaxies.  This explains the lack of star formation in these \HI\ disks: the
density of the \HI\ is simply too low.  In fact, these \HI\ disks will not
change significantly over a Hubble time.  It is still possible that this \HI\
is accreted, but this should have happened a long time ago.  Perhaps one
can consider galaxies like NGC 807 very early-type spiral galaxies, where the
gas mass of the disk is so low that only a very faint stellar disk developed. 
These galaxies could fit in the continuum of spirals to disky elliptical
proposed by e.g.\ Kormendy and Bender (1996).  Another example of such a
regular, low surface-brightness \HI\ disk in a bright elliptical is NGC 2974
(Kim et al.\ 1988).  NGC 2974 does indeed show hints of a stellar disk. 

There are \HI\ datacubes available now for about 10 low-luminosity early-type
galaxies.  In Figure 3 we show the total intensity images of two of the
objects we have observed (NGC 802 and ESO 118--G34) with ATCA.  Comparing the
\HI\ morphology with that observed in more-luminous galaxies one can make the
following two remarks:

1) The range of different \HI\ morphologies seen in luminous galaxies are not
observed in low-luminosity galaxies.  Instead, the \HI\ distribution, and
especially the kinematics, is always that of a relatively relaxed disk-like
structure, that is very extended compared to the optical image.  This does not
mean that the \HI\ in these galaxies is not accreted.  In several galaxies
there are signs that the \HI\ is in fact accreted.  For example, in NGC 802
the kinematical major axis is perpendicular to the optical major axis, clearly
suggesting that the \HI\ is accreted (see Fig.\ 4).  Another low-luminosity
elliptical showing such a mis-alignment is NGC 855 (Walsh et al.\ 1990, Knapp
priv.\ comm.).  The velocity field of the \HI\ in NGC 802 is very regular, so
the accretion must have happened some time ago.  Also the velocity field of
ESO 118--G34 (see Fig.\ 4), although still relatively regular, shows some
features indicating that the \HI\ is accreted.  Also some of the
low-luminosity galaxies observed by Lake et al.\ (1987) show similar
characteristics.  In the other two galaxies we have observed (NGC 2328 and ESO
027--G21), the kinematics of the
\HI\ is very regular and does not suggest that the \HI\ has been accreted
recently.

\begin{figure}
\centerline{\psfig{file=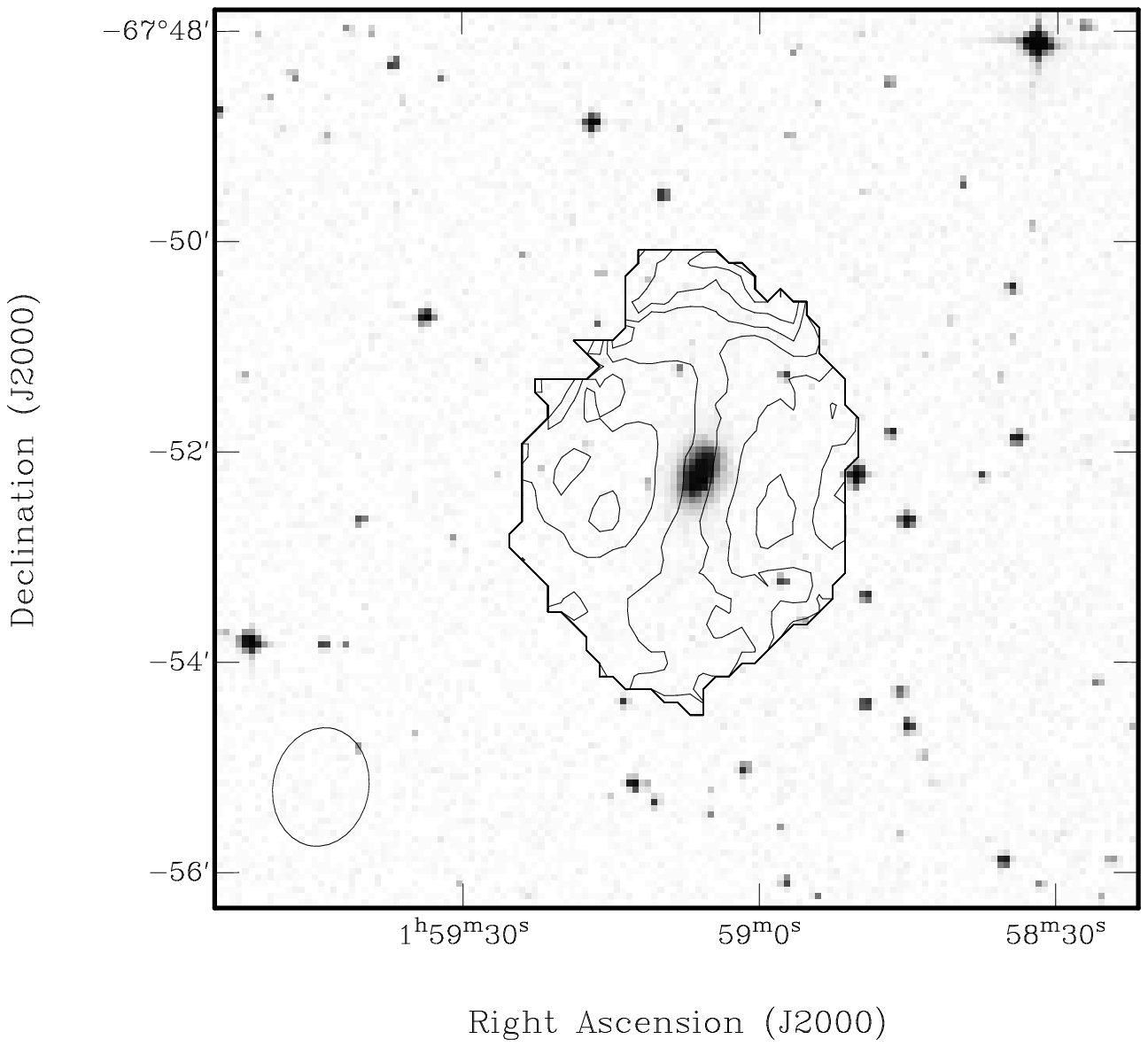,width=5cm}
\hss
\psfig{file=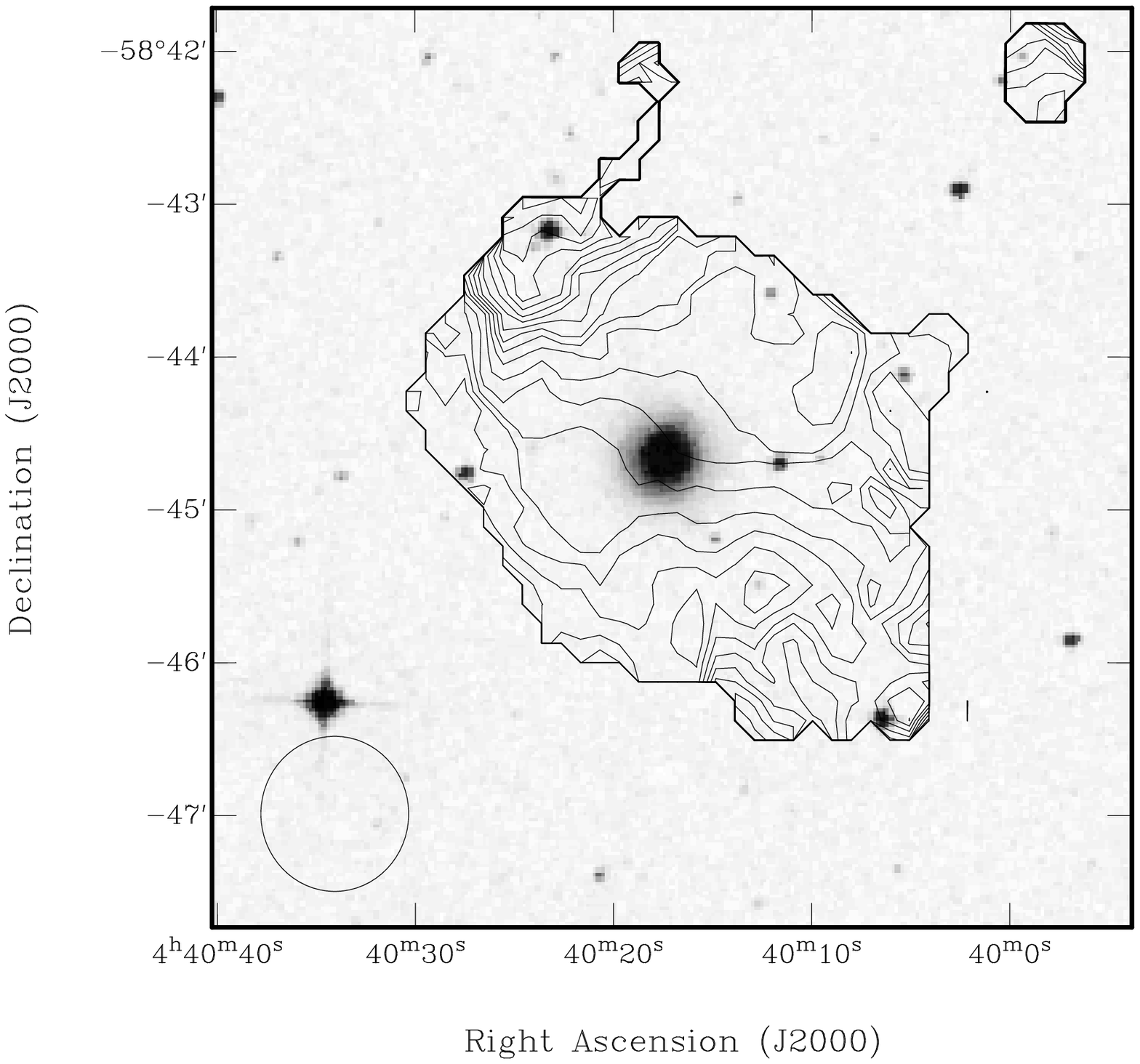,width=5cm}}

\centerline{\psfig{file=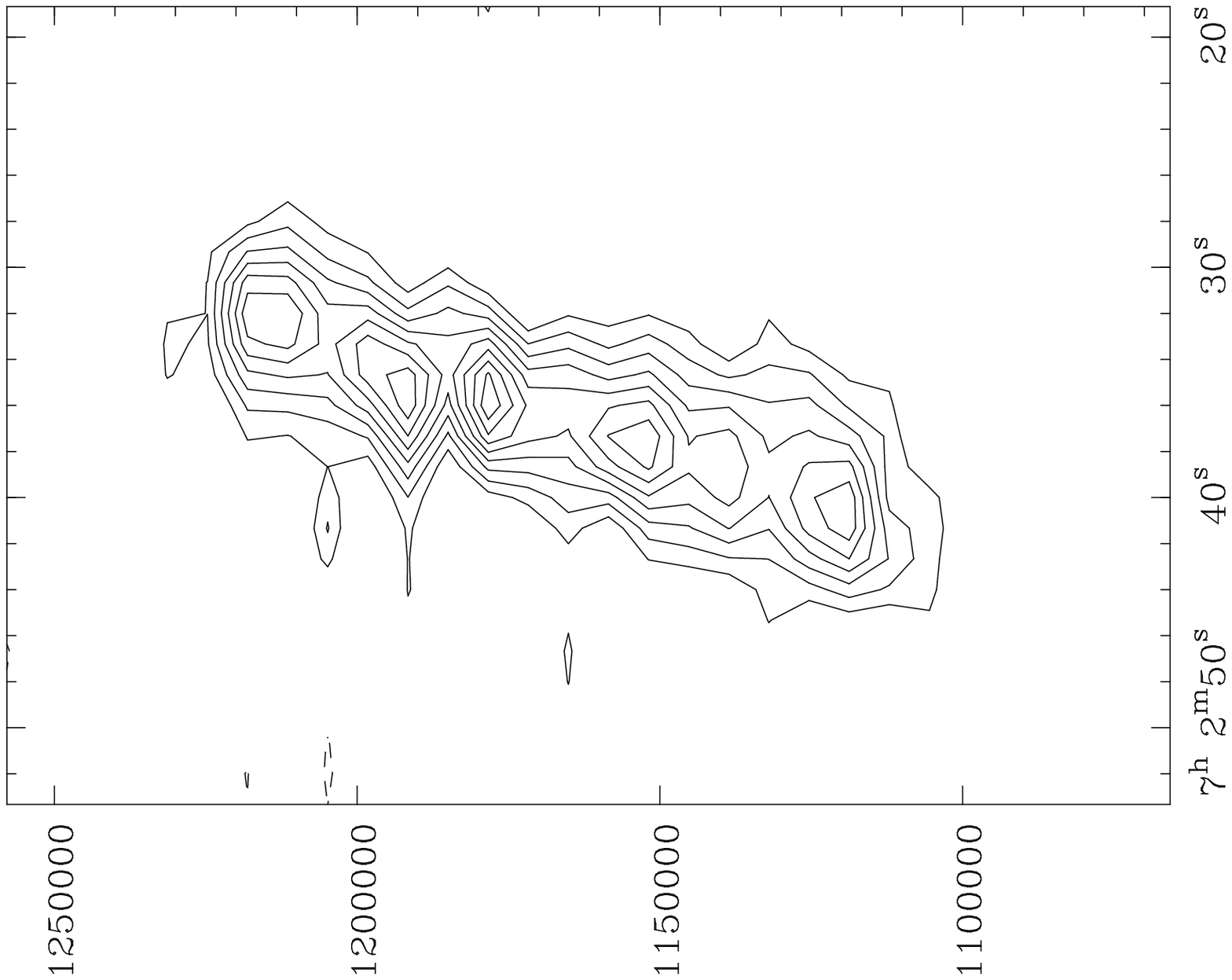,height=5cm,angle=-90}
\hss
\psfig{file=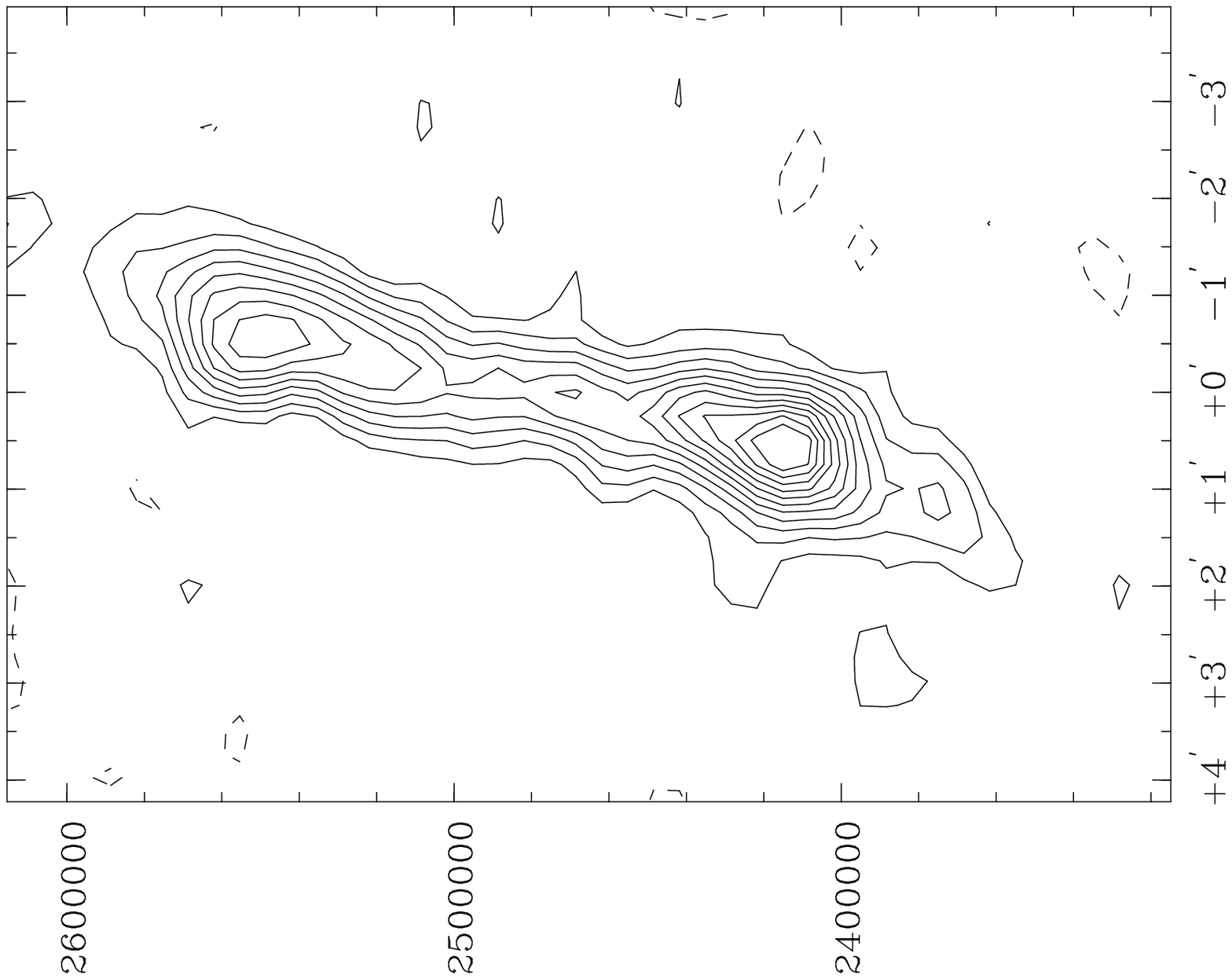,height=5cm,angle=-90}}

\caption{Velocity fields of NGC 802 {\sl (top left)} and ESO 118--G34 {\sl (top
right)}; position-velocity plots of NGC 2328 {\sl (bottom left)} and ESO
027--G21 {\sl (bottom right)}.} 

\end{figure}

In the four galaxies  we observed, star formation is  occurring in the central
10-20 arcsec.  The rate of star formation, estimated from the \Halpha\ fluxes,
is such that the timescale for consuming all \HI\ is quite short in two of the
galaxies (about 10$^9$ yr, NGC 2328, ESO 118--G34),  while in the other two it
is an order of magnitude  longer.  Therefore, in  two of the four galaxies the
star formation has to be short-lived. This is another indication that the \HI\
is accreted in at least some galaxies.

2) The \HI\  disks observed extend all  the way towards  the centre.  In fact,
the \HI\ distribution is quite peaked towards the centre, something definitely
not observed  in   more massive galaxies.   The   values of the   central \HI\
surface-brightnesses  are also  higher   compared to those  observed  in  more
luminous galaxies.  Figure 5 shows the radial \HI\ surface density profiles of
two of  the galaxies we  have observed.  These  plots show that in the centres
the density can go up to about at least 4 \Msol\,pc$^{-2}$.  Given the limited
resolution of our observations, the actual \HI\ surface density is quite likely
to be   higher  near the  centre.    These high  \HI\   surface  densities are
consistent with  the  fact that we do  observe  star formation in the  central
10-20 arcsec in all four galaxies.  However,  in most of  the galaxy, the \HI\
surface density is    below 1 \Msol\,pc$^{-2}$. So,   contrary  from the  star
formation near the centre, not much star formation will occur at larger radii.

\begin{figure}
\centerline{\psfig{file=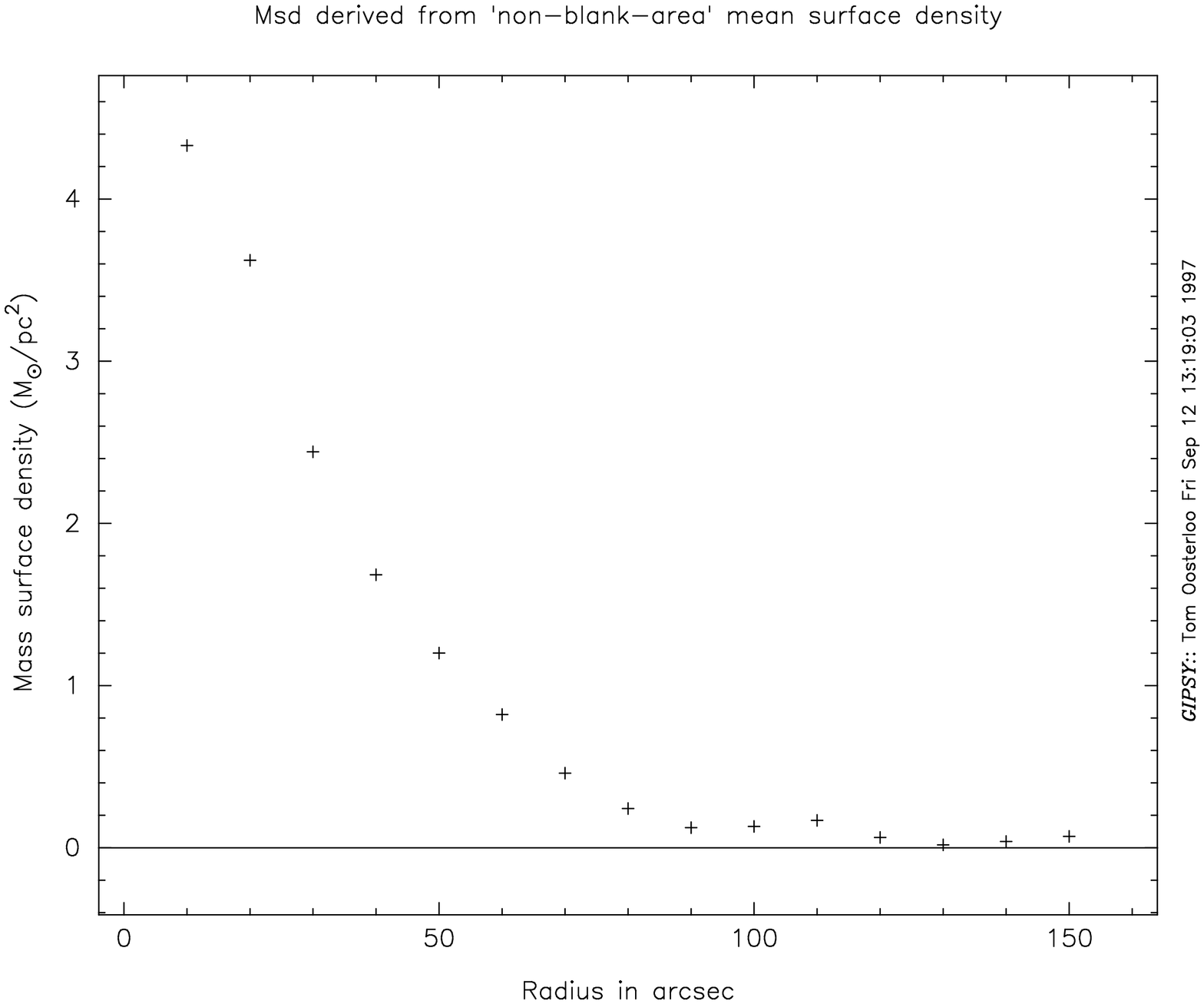,width=6.5cm,angle=-90}
\hss
\psfig{file=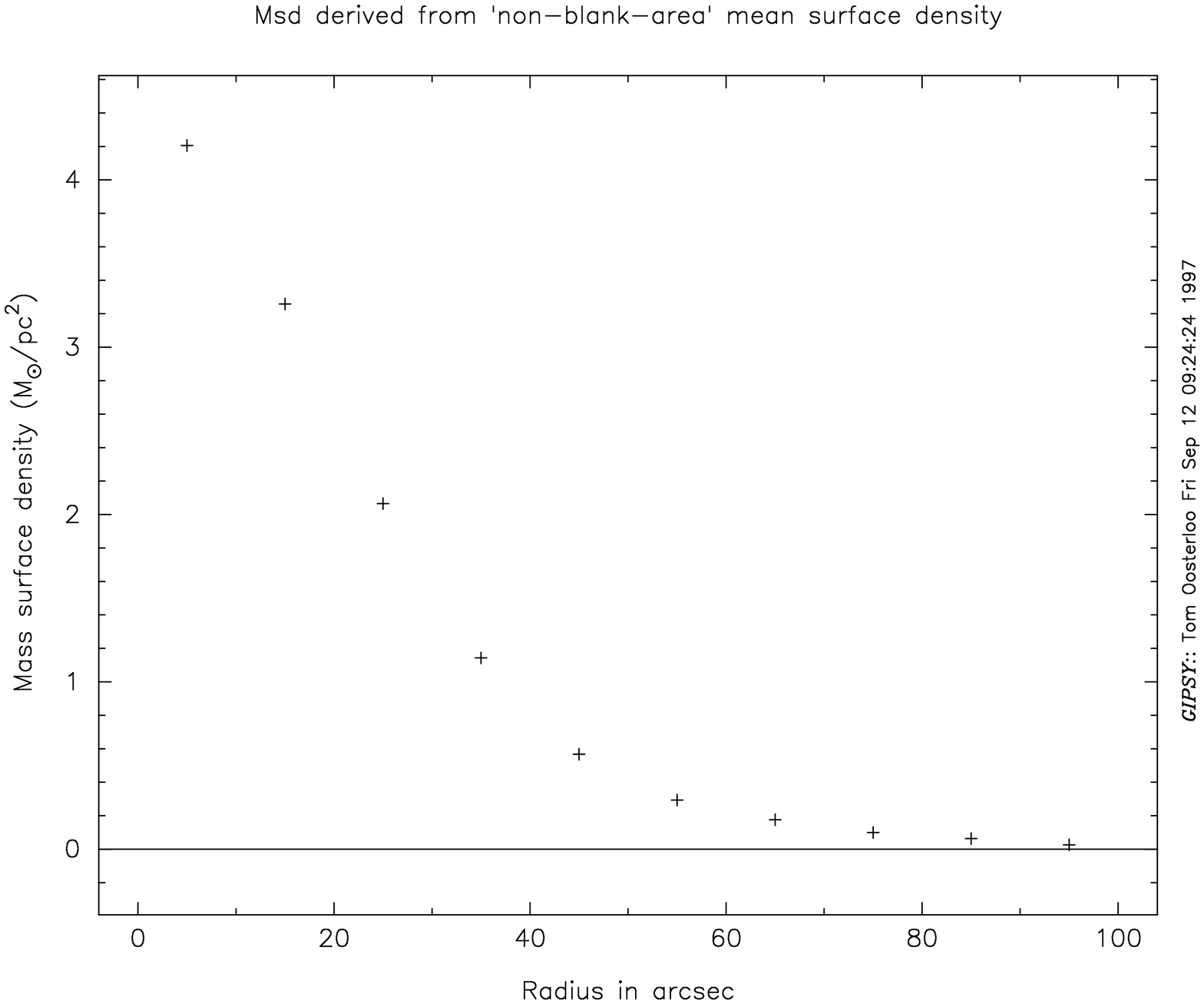,width=6.5cm,angle=-90}}
\caption{Radial \HI\ surface density profiles of NGC 802 {\sl (left)} and ESO
118--G34 {\sl (right)}}
\end{figure}

\section{Discussion}

The predominance  of   the disk-like \HI\   morphology  in  the low-luminosity
galaxies is consistent with the observation  that disky ellipticals tend to be
of  lower luminosity  and that   lower   luminosity galaxies  are more   often
rotationally  supported: we observe the  gas component  from which faint disks
can  have  formed.  It  is not entirely   clear  why low-luminosity early-type
galaxies  more often have   a regular \HI\   disk.  In  several low-luminosity
galaxies, there is still evidence that the \HI\  is accreted.  It appears that
the accretion happens in a  different way in  smaller ellipticals.  One factor
could be  the  environment.  The  luminous  ellipticals  with messy   \HI\ are
usually near the  centre of a small group,  while the low-luminosity  galaxies
observed  are more  isolated.   That these luminous  ellipticals are sometimes
near the centre of a group could be the result of the evolution of that group.
Perhaps interactions happen too often in such  a group for  the \HI\ to settle
in  a nice disk, while in  the more isolated  low-luminosity galaxies the \HI\
has time to  settle.  The more luminous ellipticals  with a regular disk (like
NGC 807) are also relatively isolated, supporting  this view.  In both IC 1459
and  in NGC 5077 several different  \HI\ systems are observed, indicating that
more than one  interaction  is responsible for  the \HI\  observed  around the
galaxies.  Another factor could be that the low-luminosity galaxies accrete at
a slower, perhaps more constant rate, from smaller companions.

That the surface  density  of the \HI\   in  low-luminosity galaxies  is  more
centrally   concentrated,  is perhaps   also  the  result of   a more  `quiet'
accretion, but it could also mean that the centres of luminous ellipticals are
more hostile to \HI.  Perhaps interaction with the stellar winds (as suggested
by Lake et al.\ 1987), or with a halo of hot X-ray gas ionises the neutral gas
in the central regions of luminous ellipticals, leaving a  disk of ionized gas
near the  centre (as observed in several   luminous elliptical galaxies).  The
different character of    the  spectrum of    the ionized  gas  in  high-  and
low-luminosity galaxies  suggests  that the   conditions are  different.  This
hostile environment would prevent a build-up of  \HI\ near the centre and star
formation will not occur in  such galaxies.  In low-luminosity galaxies  these
interactions do not occur, or are less strong, and \HI\ is found even close to
the  centre, with densities high enough  for stars to form.   This may also be
connected to the different stellar density distributions observed in high- and
low-luminosity galaxies.

\newpage

\noindent{\bf Questions}

\noindent
{\sl Trinchieri:} Is there now enough overlap between \HI- and X-ray data, so
that you could test the issue that X-ray emitting gas could provide a hostile
environment for the \HI\ in more luminous systems?

\smallskip
\noindent
{\sl Oosterloo:}  Yes,  one would  like  to see  if there is   any correlation
between  the X-ray properties and  the \HI. Unfortunately  there  is only very
little overlap. One  could  study  one  or two   galaxies,  but it  would   be
impossible to do statistics.

\smallskip
\noindent
{\sl  Pahre:}   In the  introduction by Gonzales,    he showed  how elliptical
galaxies follow different scaling relations from the dwarf ellipticals end the
dwarf spheroidals. Are the  galaxies in your  sample the low extension  of the
elliptical galaxy sequence, or are some or all dwarf ellipticals?

\smallskip
\noindent
{\sl Oosterloo:} They are all small versions of the luminous galaxies.

\smallskip
\noindent
{\sl  Hau:} The difference   between luminous and low-luminosity  ellipticals,
could it be because you are comparing apples with  oranges? Many of the things
you showed with    holes in the   centre  are merger remnant  candidates.   Do
high-luminosity Es without shells have centrally peaked profiles?

\smallskip
\noindent
{\sl Oosterloo:} Indeed,  based on the characteristics
of the  \HI,   luminous and  low-luminosity  galaxies  have  had  a  different
evolution, this is exactly the point. I don't think one can say much about the
\HI\ in luminous galaxies without shells.  The presence of \HI\ correlates very
strongly with optical peculiarities, so there is very little data on
ellipticals without shells. The only case is perhaps NGC 807, which indeed has
a centrally peaked density distribution. One should note that if there is a
central hole in the \HI, it is very often filled up with an disk of ionized
gas. So the gas disk is there, but it is in a different state.

\smallskip
\noindent
{\sl Goudfrooij:}   Do you know   of any   spectral  data  of   low-luminosity
ellipticals containing \HII\ that can  be used to  derive the metallicities of
the ionized gas?

\smallskip
\noindent
{\sl  Oosterloo:} I am not  aware  of any  such data.  It  would certainly  be
worthwhile obtaining such data.

\smallskip
\noindent
{\sl Andernach:} Do you know the kinematics from optical spectroscopy for the
regular \HI-disk-low-luminosity galaxies? 

\smallskip
\noindent
{\sl Oosterloo:} Also for this there is no data  available, we hope to be able
to obtain such data soon.

\smallskip
\noindent
{\sl Eskridge:} For NGC 802 you show  that the \HI\  rotates about the optical
major axis. Do you know what the optical kinematics are?

\smallskip
\noindent
{\sl  Oosterloo:}  No, we do   not  know what the    kinematics of the stellar
component is in this galaxy. It would be interesting to obtain optical spectra
of NGC 802 and see if this is a polar ring galaxy.

\smallskip
\noindent
{\sl Avila-Reese:}  1)  Is it possible to   measure the thickness of the  \HI\
disks in the early-type galaxies?   2) Could the  surface density profiles you
have shown  for   the \HI\ disks  be  fitted  with an exponential  law?  Which
scalelengths have these disks?

\smallskip
\noindent
{\sl Oosterloo:} 1) In  principle this is   possible, although in practise  it
would   be very difficult   to  get data with   the  required S/N and  spatial
resolution.  2) I haven't fitted  any model to  the  \HI\ distributions, so  I
cannot really answer  this question. If  one would  fit a  single exponential,
given the central concentration of the  \HI, the scalelength would probably be
quite small.

\end{document}